\documentclass[prl,twocolumn,showpacs]{revtex4}

\usepackage[dvips]{graphicx}

\usepackage{amsfonts}
\usepackage{amssymb}
\usepackage{amsmath}
\usepackage{bbm}
\usepackage{graphicx}

\newcommand{\ket}[1]{\text{$|#1\rangle$}}
\newcommand{\bra}[1]{\text{$\langle#1|$}}
\newcommand{\vac}{\ket{\text{vac}}}
\newcommand{\sqv}{\text{$\ket{\text{vac}(\eta)}_{\text{sq}}$}}

\newcommand{\one}{\mbox{$1 \hspace{-1.0mm}  {\bf l}$}}

\begin{document}

\title{Observable geometric phase induced by a cyclically evolving dissipative process}
\author{Angelo Carollo }
\affiliation{Institute of Quantum Optics and Quantum Information, Technikerstraße 21a, A-6020 Innsbruck, Austria}
\author{G. Massimo Palma}
\affiliation{NEST - CNR \& Dipartimento  di Scienze Fisiche ed Astronomiche, Universit\'a di Palermo, via Archirafi 36, I-90123
Palermo, Italy}

\begin{abstract}
In \cite{CP}  a new way to generate an observable geometric phase on a quantum system by means of a completely incoherent phenomenon has been proposed. The basic idea is to force the ground state of the system to evolve ciclically  by "adiabatically" manipulating the environment with which it interacts. The specific
scheme analyzed in \cite{CP}, consisting of a multilevel atom interacting with a broad-band squeezed vacuum bosonic bath whose  squeezing parameters are
smoothly changed in time along a closed loop, is here solved in a more direct way. This new solution emphasizes how the geometric phase on the ground state of the system is indeed due to a purely incoherent dynamics.
\end{abstract}


\maketitle

\section{Introduction}
Geometric phases appear when a quantum state undergoes a cyclic evolution. The most celebrated example of geometric phase is perhaps the Berry phase \cite{Berry}. Let us illustrate the phenomenon with a simple system: a spin in a magnetic field whose direction changes slowly in time. The Hamiltonian of such system is

\begin{equation}
H = \frac{1}{2} {\bf B}\cdot{\boldsymbol \sigma}
\end{equation}
where $\boldsymbol{\sigma} \equiv (\sigma_x,\sigma_y,\sigma_z)$ are Pauli operators and ${\bf B}(t)\equiv B_0 ( \sin\vartheta \cos\varphi , \sin\vartheta \sin\varphi , \cos\vartheta )$ is a three dimensional vector, which we assume to be time dependent. The energy eigestates of $H_S$ are  the eigenstates of the operator $ {\boldsymbol \sigma}\cdot{\bf n}$ where $\bf n\equiv (\sin\vartheta \cos\varphi ,
\sin\vartheta \sin\varphi , \cos\vartheta )$ is a unit vector pointing in the instantaneous ${\bf B}$ direction and can be written as

\begin{eqnarray}
|\uparrow_n \rangle =  \cos \frac{\vartheta }{2}|\uparrow_z \rangle
+ e^{i\varphi}\sin \frac{\vartheta }{2}|\downarrow_z \rangle\nonumber\\
|\downarrow_n \rangle = \sin \frac{\vartheta}{2}|\uparrow_z \rangle- e^{i\varphi}\cos \frac{\vartheta }{2}|
\downarrow_z\rangle
\end{eqnarray}
where $\ket{\downarrow_z} , \ket{\uparrow_z} $  are the eigenstates of the operator $\sigma_z$.
Let us assume that  $\bf n$ changes slowly in time so that the so called adiabatic approximation is valid and that at $t=0$ the system is prepared in an energy eingenstate, say $\ket{\psi (0))} =  \ket{\uparrow_n (0) }$. At  at time $t$  we have
$\ket{\psi (t))} =  \ket{\uparrow_n (t) }$ i.e. the state of the system adiabatically follows the direction of $\bf n$.  Let us now further assume that  ${\bf B}(T) = {\bf B}(0)$ i.e.  the adiabatic motion of ${\bf B}$ is cyclical. In this case the energy eigenstrates acquire, on top of the dynamical phase, a  phase factor $\chi_g$ that depends only on the geometry of the path followed by ${\bf B}(t)$. In mathematical terms
\begin{equation}
\ket{\uparrow_n (T)} = e^{i\chi_g}e^{iB_0T} \ket{\uparrow_n (0)}
\end{equation}
where for simplicity $\hbar =1$. The geometric phase is given, according to Berry \cite{Berry}, by
\begin{equation}
\chi_g=i\oint \bra {\uparrow_n} d \ket{\uparrow_n}
\end{equation}
A straightforward calculation shows that $\chi_g$ is equal to the solid angle spanned by ${\bf B}(t)$ in his cyclic evolution and it is therefore independent on $T$. Note, by the way, that the eigenstates $\ket{\downarrow}_n$ and $\ket{\uparrow}_n$ acquire a geometric phase of opposite sign.

The role of the adiabatic approximation in the example above is to provide a way to perform a parallel transport of a state vector along a suitable path in Hilbert space by slowly changing a time dependent system hamiltonian. One may wonder if the same goal can be achieved by changing slowly in time the irreversible dynamics of a system coupled with a reservoir. To be more specific assume that a system interacting with a rigged reservoir relaxes to an equilibrium pure state $\ket{\psi_s}$. Such state will in general depend on the properties of the reservoir. If the properties of the reservoir are changed slowly in time the equilibrium state $\ket{\psi_s}$ follows adiabatically and, if such change is cyclical, it acquires a geometric phase $\chi_g$.  Furthermore, being induced by the irreversible dynamics itself $\chi_g$ should be intrinsically immune from noise. We are not referring here to the standard resistance of geometric phase to environmental noise which has attracted much attention for its potential benefits in the implementation of quantum gates \cite{noise}. Here something conceptually different happens: it is the environment itself which generates the the geometric phase.

\section{A four level system interacting with a broadband squeezed vacuum}
The system which we will explicitly analyze here to illustrate the above idea is the same studied in \cite{CP}: a multilevel atom interacting with a broadband squeezed vacuum. The interesting feature of such system is that its irreversible dynamics relaxes to a nontrivial  pure ground state, whose
configuration depends on the squeezing parameters of the reservoir.  By changing the squeezing parameters it is possible therefore possible to manipulate indirectly the ground state of the system.

For the sake of completeness let us first review the essential features of the dissipative dynamics of an atomic system interacting with a broad band squeezed vacuum reservoir. Consider first a three level atom whose interaction with an electromagnetic field in the rotating wave approximation is described by the following Hamiltonian:
\begin{equation}
         H = H_S +\int \hat{a}^{\dagger}(\omega)\hat{a}(\omega)\omega d\omega
         + \int \left[ g(\omega)S^{\dagger}\hat{a}(\omega) + H.c.
         \right] d\omega \text{,}
\end{equation}
where $H_S = \Omega\sum_{k=-1}^1k\ket{k}\bra{k}$ is the internal Hamiltonian of the atom, $S = \ket{-1}\bra{0} + \ket{0}\bra{1}$
is the atomic operator describing the absorption of an excitation by the atom and $\hat{a}(\omega)$ is the annihilation operator
of the mode with the energy $\omega$.  The field, which we treat as a reservoir, is assumed to be in a broad band squeezed vacuum state defined as
\begin{equation}
         \sqv = \hat{K}(\eta)\vac \text{,}
\end{equation}
where $\hat{K}(\eta)$ is a unitary multimode squeezing transformation~\cite{EkertPBK89,AgarwalP90}, which correlates symmetrical pairs of
modes around the carrier frequency $\Omega$
\begin{equation}\label{eq:squeezer}
         \hat{K}(\eta)= \exp \frac{1}{2}\left\{\int \left[\eta \hat{a}^\dag(\Omega-\omega)\hat{a}^\dag(\Omega+\omega) - H.c.\right] d\omega\right\}
       \text{,}
\end{equation}
In the above expression $\eta = e^{i\varphi}r$ is the squeezing parameter, whose polar coordinates $\varphi\in\{0\dots 2\pi\}$ and
$r>0$ are care called phase and amplitude of the squeezing, respectively. For a  broad band squeezed vacuum, characterized by a
value of squeezing parameters basically constant over a broad band, one has the following expectation value for the field bosonic
operators:

\begin{align}
& \langle a^\dag (\omega ) a (\omega ') \rangle =  \sinh^2 r \delta (\omega - \omega')\\
& \langle a (\omega ) a ^\dag(\omega ') \rangle = ( \sinh^2 r+1) \delta (\omega - \omega')\\
& \langle a (\omega ) a (\omega ') \rangle =( e^{i\varphi} \sinh r \cosh r )\delta (\omega + \omega' - 2\Omega)\\
& \langle a^\dag (\omega ) a^\dag (\omega ') \rangle =   (e^{-i\varphi} \sinh r \cosh r) \delta (\omega + \omega' - 2\Omega)
\end{align}

The use of the Born Markov approximation, justified by the broadband nature of the field, leads to a master equation
for the atomic degrees of freedom which can be written in the following compact form
\begin{equation}
\label{MasterEq2} \frac{d\rho}{dt} =-\frac{\gamma}{2}\{R^\dag R
\rho+\rho R^\dag R-2 R \rho R^\dag\}
\end{equation}
whith
\begin{equation}
\gamma = 2\pi | g(\Omega )|^2
\end{equation}
and where we have defined the new "dressed" atomic operator
\begin{equation}
\label{Rmatrix} R(\eta)=S \cosh r+e^{i\varphi}S^\dag \sinh r.
\end{equation}
From (\ref{Rmatrix}) it follows that the state
\begin{equation}
\label{eq:DFstate} \ket{\psi_{DF}(\eta)}=c \ket{-1}-e^{i\varphi}s
\ket{1},
\end{equation}
with $ c(r) = \frac{ \cosh r}{\sqrt{ \cosh 2 r }} $ and $s(r)=
\frac{ \sinh r}{ \sqrt{ \cosh 2 r }}$, satisfies the condition
$R(\eta)\ket{\psi_{DF}(\eta)}=0$. In other words, this state, being
unaffected by the environment,  is decoherence free \cite{PalmaSE96,DuanG97,ZanardiR97,LidarCW97}.
Moreover $\ket{\psi(\eta)}$ represents the new ground state, as
all the other states of the atomic system relax to it.

\begin{figure}[htb]
         \begin{center}
                 \includegraphics[width=\columnwidth]{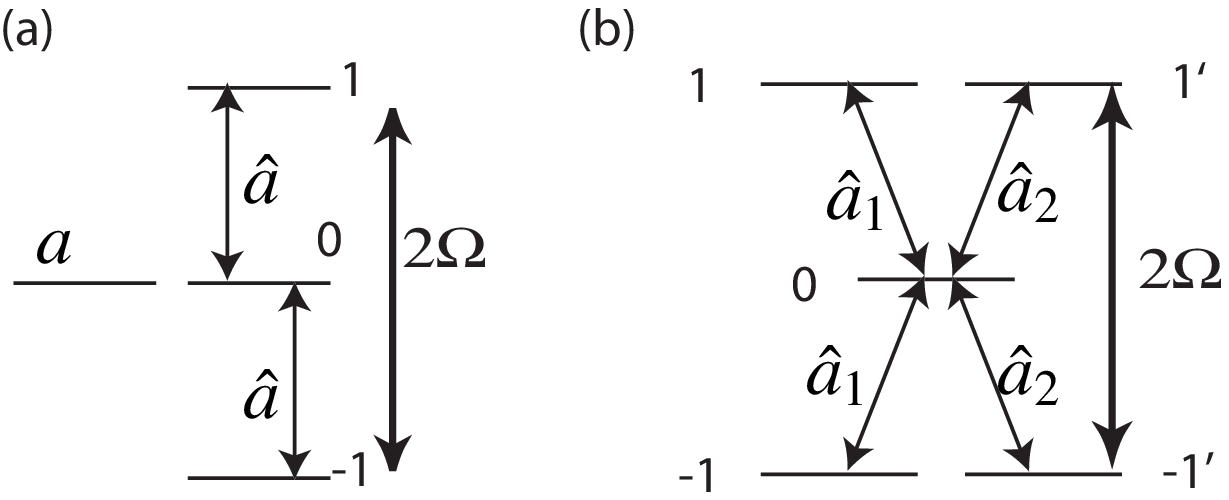}
         \end{center}
         \caption{Schematic representation of the energy levels of the two systems considered. \\
         (a) A four level systems with an equal  energy gap $\Omega$ between states $|-1\rangle$ and
         $|0\rangle$ and between  $|0\rangle$ and
         $|1\rangle$. The transitions between these level are coupled to the modes $\hat{a}(\omega)$ of the reservoir.
         The reference state $\ket{a}$ is decoupled from the reservoir.\\
         (b) A five-level system where  transitions $1\leftrightarrow 0$ and $0\leftrightarrow -1$ are
         coupled to modes $\hat{a}_1(\omega)$ while $1'\leftrightarrow 0$ and $0\leftrightarrow -1'$ are coupled to modes $\hat{a}_2(\omega)$ of the reservoir.}
         \label{fig:sys4l}
\end{figure}

So far we have assumed the squeezing parameter $\eta$ is time independent, however one may think of a scenario in which $\eta$ is slowly changed in time. In such case the reduced atomic dynamics is described by a time dependent master equation:
\begin{equation}
\label{TimeDepME} \frac{d\rho}{dt} =-\frac{\gamma}{2}\{R^\dag(t) R(t)
\rho+\rho R^\dag(t) R(t)-2 R(t) \rho R^\dag(t)\}
\end{equation}
It is reasonable to expect that if such change is made slowly enough $\ket{\psi_{DF}(\eta (t))}$ is adiabatically changed accordingly and and, if $\eta (0) = \eta (T)$, acquires a purely geometric phase. To show that this is indeed the case we consider here, with no loss of generality,  the simple case in which the squeezing amplitude $r$
 is kept constant while the squeezing phase $\varphi$  is slowly changed  from
$0$ to $2\pi$.  We show that, provided $\dot{\varphi}$ is small enough, the state $\ket{\psi_{DF}}$ is adiabatically decoupled
from its orthogonal subspace and acquires, after a cyclic evolution, a geometric phase depending, in this case, only on the amount
of squeezing $r$. Note however that, since the steering process is essentially incoherent, any phase information acquired by a
superposition of $\ket{\psi_{DF}}$ and a state belonging to the orthogonal subspace is inevitably lost, as the latter is subject
to decoherence. The only way to retain such information is to consider an auxiliary level $\ket{a}$, unaffected by the noise,
playing the role of a reference state in an interferometric measurement (see Fig.\ref{fig:sys4l}a)

 For simplicity
assume that $\ket{a}$ is unaffected by the environment during the whole evolution, and, hence, is time independent. The whole
information about the geometrical phase and the coherence retained by the system during its evolution is then recorded in the
phase and amplitude of the density matrix term $\bra{a}\rho\ket{\psi_{DF}}$. In \cite{CP} the system dynamics was analyzed in a
suitable "rotating" frame. Although mathematically more elegant such approach may, at first glance, mask the fact that the system
dynamics is fully incoherent. Here therefore we will explicitly solve (\ref{TimeDepME}) with no change of frame of reference. We
must therefore calculate

\begin{align}\label{eq1}
\frac{d}{dt}\bra{a}\rho\ket{\psi_{DF}} & = \bra{a}\dot{\rho}\ket{\psi_{DF}} + \bra{a}\rho\frac{d}{dt}\ket{\psi_{DF}}\\
& = -i\dot{\varphi}s e^{i\varphi}\bra{a}\rho\ket{1}\nonumber\\
&= -i\dot{\varphi}s \left[ c \bra{a}\rho\ket{\psi_{\bot}}-s\bra{a}\rho\ket{\psi_{DF}}\right]\nonumber
\end{align}
where we have made use of the fact that  $R\ket{\psi_{DF}}=0$ and $R \ket{a} = R^{\dagger}\ket{a} = 0$ and made the substitution

\begin{equation}\label{psiort}
\ket{1}=e^{-i\varphi}\left[ c\ket{\psi_{\bot}}-s\ket{\psi_{DF}}\right]
\end{equation}
with
\begin{equation}\label{ortog}
\ket{\psi_{\bot}}=s\ket{-1}+e^{i\varphi}c\ket{1}.
\end{equation}
The linear system can be closed with the following equation of motion
\begin{align}
\frac{d}{dt}\bra{a}\rho\ket{\psi_{\bot}} \label{eq2}&=\bra{a}\dot{\rho}\ket{\psi_{\bot}}+\bra{a}\rho\frac{d}{dt}\ket{\psi_{\bot}}\\
&=-\frac{\tilde{\gamma}}{2}\bra{a}\rho\ket{\psi_{\bot}} +i\dot{\varphi}c e^{i\varphi}\bra{a}\rho\ket{1}\nonumber\\
&=-\frac{\tilde{\gamma}}{2}\bra{a}\rho\ket{\psi_{\bot}}+ i\dot{\varphi}c \left[ c \bra{a}\rho\ket{\psi_{\bot}}-s\bra{a}\rho\ket{\psi_{DF}}\right]\nonumber
\end{align}
where $\tilde{\gamma}=\gamma\cosh 2r$ and we have made use of the fact that   $R^\dag R\ket{\psi_{\bot}}=\cosh 2r
\ket{\psi_{\bot}}$ and $R \ket{a} = R^{\dagger}\ket{a} = 0$. Eq. (\ref{eq1},\ref{eq2}) therefore form the following linear system
\[
\frac{d\mathbf{v}}{dt}=-iG\cdot \mathbf{v}
\]
where
\[
\mathbf{v}=\left(\begin{array}{c} \bra{a}\rho\ket{\psi_{DF}}\\\bra{a}\rho\ket{\psi_{\bot}}\end{array}\right)
\]
and
\begin{equation}\label{G}
G=\left(\begin{array}{cc} -\dot{\varphi} s^2 &\dot{\varphi} s c \\ \dot{\varphi} s c &-\dot{\varphi} c^2 -i\tilde{\gamma}/2
\end{array}\right)
\end{equation}

Although such system can be easily solved diagonalising $G$, it is however clear that, in the limit $\tilde{\gamma}\gg \dot{\varphi}$ the off-diagonal terms are exponentially suppressed by the incoherent term. Hence, the leading contribution to the time evolution of the initial $\ket{\psi_{DF}}$ state is a phase factor $e^{i\chi_g}$, with
\begin{equation}
\chi_g=2\pi s^2 = 2\pi \frac{\sinh^2r}{\cosh 2r}.
\label{chi}
\end{equation}
This demonstrate that,  in the lowest order
"adiabatic" approximation, the system preserves its coherence. In
fact the amplitude damping
of $\rho_{\psi a}$ occurs only when we take into account the first
order contribution in  $\dot{\varphi} / \gamma$.  It follows
that for small $\dot{\varphi}$ the system admits an adiabatic
limit, in which the subspace $H_{DF}(t)$ spanned by
$\ket{\psi_{DF}(t)}$ and $\ket{a}$ is adiabatically decoupled from
its orthogonal subspace ${\cal H}_{\bot}(t)$. For this reason,
${\cal H}_{DF}(t)$ is decoupled from the effects of the decoherence,
which only affect states lying in its orthogonal counterpart.
Note that this adiabatic limit is different from the "standard" adiabatic approximation, where the timescale is set by the energy splitting $\Omega$ between energy levels. Indeed in order to observe a geometric phase generated by a slowly varying Hamiltonian in the presence of damping one must satisfy the condition $\Omega \gg \dot{\varphi} \gg \gamma$. In our scheme, as the phase is generated by a slowly varying limblandian, one must instead satisfy $\dot{\varphi} \ll \gamma$. Within this approximation, then, a state prepared in the space
${\cal H}_{DF}(0)$ is adiabatically transported rigidly inside the
evolving subspace ${\cal H}_{DF}(t)$. As a result of this adiabatic
steering, when the system is brought back to its initial
configuration, the coherence $\rho_{\psi a}$ acquires a phase
$\chi_g$ that can be interpreted as the
geometric phase accumulated by the state $\ket{\psi_{DF}(t)}$.
Note indeed that according to the canonical formula for the Berry phase, the geometric phase of $\ket{\psi_{DF}(t)}$ is given by
\begin{eqnarray*}
\label{eq:geomphase}
\chi_g&=&i\oint \bra{\psi_{DF}}d\ket{\psi_{DF}}=\\
&=&i\int_0^{2\pi}
\bra{\psi_{DF}}\frac{d}{d\varphi}\ket{\psi_{DF}}d\varphi= 2\pi s^2\text{.}
\end{eqnarray*}
which coincides with (\ref{chi}).
As anticipated the value of $\chi_g$ depends only on the squeezing,
and vanishes as the squeezing tends to zero. Moreover, notice that
the phase $\chi_g$ is purely geometrical, i.e. there is not any
dynamical term generated by a Hamiltonian term, as,
in absence of a streering process, the states inside $H_{DF}$ have
a trivial dynamics. This feature makes the measurement of
this phase a relatively easy task. Indeed the absence of a dynamics makes unnecessary the use of spin-echo techniques in an interferometric experiment.

\section{A five level system}
To conclude let us re-analyse the detection scheme proposed in \cite{CP}.
 Let's consider the five-level system as shown in
picture~(\ref{fig:sys4l}). This system consists of two
replicas of the three-level system discussed so far, with the
level $\ket{0}$ in common. The important ingredient is that
transitions $\ket{0}\leftrightarrow \ket{1}$ and
$\ket{-1}\leftrightarrow \ket{0}$ are coupled to a set of field modes different from those inducing transitions
$\ket{0}\leftrightarrow \ket{1'}$ and $\ket{-1'}\leftrightarrow
\ket{0}$. A simple way to achieve this, is to choose, for example,
polarisation selective transitions, say, left-circularly polarised
modes for the former transitions and right-circularly polarised
for the latter ones. The Hamiltonian of such system is
\begin{eqnarray}\label{Hamiltonian2}
         H &=& H_S +\sum_{i=1,2}\int \hat{a}^{\dagger}_i(\omega)\hat{a}_i(\omega)\omega d\omega\\
         &&+ \sum_{i=1,2}\int \left[ g_i(\omega)S_i^{\dagger}\hat{a_i}(\omega) + H.c.
         \right] d\omega \text{,}\nonumber
\end{eqnarray}
where $H_S
=\Omega\sum_{k=-1}^1k\left(\ket{k}\bra{k}+\ket{k'}\bra{k'}\right)$,
and $S_1 =\ket{-1}\bra{0} + \ket{0}\bra{1}$ and $S_2
=\ket{-1'}\bra{0} + \ket{0}\bra{1'}$, and $\hat{a_i}(\omega)$ is
the annihilation operator of the mode with the energy $\omega$ and
polarisation $i=\{1,2\}$. Assume that both set of modes  $\hat{a}_1(\omega)$ and $\hat{a}_2(\omega)$ are in a broad band squeezed vacuum, although with different squeezing parameters $\eta_1=r_1e^{i\varphi_1}$
and $\eta_2=r_2e^{i\varphi_2}$:
\begin{equation}
\label{MEfive}
 \ket{\text{vac}(\eta_1,\eta_2)}_{sq} = \hat{K_1}(\eta_1)\hat{K_2}(\eta_2)\vac \text{,}
\end{equation}
 where $\hat{K_i}(\eta_i)$ are the analogous of the operator~(\ref{eq:squeezer}) acting on modes $\hat{a}_i$.
 Under the same assumptions described in the previous section we obtain a master equation which can be cast in the following compact form
\begin{equation}
\label{MasterEq2} \frac{d\rho}{dt}
=-\sum_i\frac{\gamma_i}{2}\{R_i^\dag R_i \rho+\rho R_i^\dag R_i-2
R_i \rho R_i^\dag\}
\end{equation}
where
\[\gamma_i = 2\pi | g_i(\Omega )|^2
\]
and
\[R_i(\eta_i)=S_i \cosh r_i+e^{i\varphi_i}S_i^\dag \sinh r_i
\]
Such  system admits a
two-dimensional decoherent-free subspace, spanned by states
$\ket{\psi_1}$ and $\ket{\psi_2}$ defined in the analogous way as
$\ket{\psi_{DF}}$ of equation~(\ref{eq:DFstate}), where $c$ and $s$ are replaced by the corresponding $c_i=\cosh r_i/\sqrt{\cosh 2r_1}$ and $s_i=\sinh r_i/\sqrt{\cosh 2r_i}$, $i=1,2$ .

Again we assume time dependent squeezing parameters
$\eta_i^t$. In particular let us assume constant $r_1 \neq r_2$ and ${\dot \varphi_1} = {\dot \varphi_2}$.

Suppose that
the system is initially prepared in a coherent superposition of
state $\ket{\psi_1}(\eta_1^0)$ and $\ket{\psi_2(\eta_2^0)}$, for
example:
\begin{equation}
\label{eq:psi1psi2}
\ket{\psi(0)}=\frac{1}{\sqrt{2}}\left(\ket{\psi_1(\eta_1^0)}+\ket{\psi_2(\eta_2^0)}\right).
\end{equation}
On the light of the previous consideration, it is expected that, when the parameters $\varphi_1, \varphi_2$ close a loop, at time
$t=T=2\pi/\dot{\varphi}$, the coherence has gained a phase
which is the difference between the expected geometric phases
$\chi_{g1} - \chi_{g2}$ acquired by the states $\ket{\psi_1}, \ket{\psi_2}$,
respectively.
To prove this we should calculate the time dependence of $\bra{\psi_1}\rho\ket{\psi_2}$. To this end, we follow the analogous procedure of the previous section, which yields to the equation:
 \begin{widetext}
\begin{align*}
\frac{d\bra{\psi_1}\rho\ket{\psi_1}}{dt}& =
\bra{\psi_1} \dot{\rho} \ket{\psi_2} + \bra{\dot{\psi}_1} \rho \ket{\psi_2} + \bra{\psi_1} \rho \ket{\dot{\psi}_2}\\ & =-i\dot{\varphi} \left[ (s_1^2-s_2^2) \bra{\psi_1}\rho\ket{\psi_2}-s_1c_1 \bra{\psi_1^{\bot}}\rho\ket{\psi_2}+ s_2c_2\bra{\psi_1}\rho\ket{\psi_2^{\bot}}\right]
\nonumber
\end{align*}
\end{widetext}
where $\psi_i^{\bot}$ are defined in analogy to Eq.~(\ref{ortog}). This equation has been obtained by inserting the master equation~(\ref{MEfive}) and using the following properties: $R_{1,2}\ket{\psi_{1,2}}=0$, and $R_{i\ne j}\ket{\psi_j^\bot}=0$.
To close the equation of motion, we need to calculate the derivative of the other matrix density elements. Eventually, making use of the fact that $R^\dag_{i\ne j}\ket{\psi_j^\bot}=R^\dag_{i\ne j}\ket{\psi_j}=0$ and $R^\dag_{i}R_i\ket{\psi_i^\bot}= \cosh 2r_i \ket{\psi_i^\bot}$ leads to a system of equation, which can be expressed in the following compact form
\[
\frac{d\mathbf{v}}{dt}=-iK\cdot \mathbf{v}
\]
where
\[
\mathbf{v}=\left(\begin{array}{c} \bra{\psi_1}\rho\ket{\psi_2}\\\bra{\psi_1}\rho\ket{\psi_2^{\bot}}\\\bra{\psi_1^{\bot}}\rho\ket{\psi_2}\\\bra{\psi_1^{\bot}}\rho\ket{\psi_2^{\bot}}\end{array}\right)
\]
and where
\begin{widetext}
\begin{equation}
K=\dot{\varphi}\times
\left(\begin{array}{cccc} s_1^2\! \!-\!\!s_2^2 &  s_2 c_2 & -  s_1 c_1& 0 \\ s_2 c_2 & s_1^2\! \!-\!\!c_2^2 -i\frac{\tilde{\gamma}_2}{2\dot{\varphi}}&0&- s_1c_1\\
- s_1c_1 &0& c_1^2\! \!-\!\!s_2^2-i\frac{\tilde{\gamma}_2}{2\dot{\varphi}}& s_2 c_2\\
0&- s_1 c_1& s_2 c_2& c_1^2\! \!-\!\!c_2^2-i\frac{\tilde{\gamma}_1+\tilde{\gamma}_2}{2\dot{\varphi}}
\end{array}\right);
\end{equation}
\end{widetext}
with $\tilde{\gamma}_i=\cosh 2r_i \gamma$.
This apparently complicated expression, can be cast in a tensor product structure, much easier to interpret. Infact $K=\one_1 \otimes G_2+G_1\otimes \one_2$, where $G_i$ are $2\times2$ matrices analogous to (\ref{G}):
\[
G_1=\left(\begin{array}{cc} \dot{\varphi} s_1^2 &-\dot{\varphi} s_1 c_1 \\ -\dot{\varphi} s_1 c_1 &\dot{\varphi} c_1^2 -i\tilde{\gamma}_1/2
\end{array}\right),
\]
and
\[
G_2=\left(\begin{array}{cc} -\dot{\varphi} s_2^2 &\dot{\varphi} s_2 c_2 \\ \dot{\varphi} s_2 c_2 &-\dot{\varphi} c_2^2 -i\tilde{\gamma}_2/2
\end{array}\right).
\]

We can thus apply step by step the reasoning of the previous section, and infer the evolution in the adiabatic limit
$\tilde{\gamma}_i\gg \dot{\varphi}$. In this limit, the coherence $\bra{\psi_1}\rho\ket{\psi_2}$ is not reduces in module, and
acquires a phase $\chi_g$ which is the difference of the geometric phase acquired by $\phi_1$ and $\psi_2$:

\[
\chi_g=\chi_1-\chi_2 \qquad \text{with } \quad \chi_i=2\pi \frac{\sinh^2 r }{\cosh 2r}
\]

As in the previous scheme, the visibility is only reduced by a factor which is linear in the ``adiabatic parameters''
$\dot{\varphi}/\gamma_i$, which guarantees the existence of the adiabatic limit. The advantage of this modified scheme is that the
value of the geometric phases can be readily measured from the polarisation of the light emitted when the system relaxes. Infact,
if the value of the squeezing parameters are suddenly changed from $r_i$ to zero, the states $\ket{\psi_i}$ are no longer
decoherent free, and and decay to the the ground states $\ket{-1}$ and $\ket{-1'}$. Such dissipative process is accompanied by two
photon emissions into the reservoir. Due to the structure of the interaction~(\ref{Hamiltonian2}) with the reservoir, the photons
emitted due to the transitions $\ket{1}\to\ket{0}$ and $\ket{1'}\to\ket{0}$, are polarised according to the geometric phase
accumulated between $\ket{\psi_1}$ and $\ket{\psi_2}$. For example, if $\hat{a}_1(\omega)$ and $\hat{a}_2(\omega)$ are right and
left circularly polarised modes, respectively, the first emitted  photon will be linearly polarized along a polarization plane
depending on $ \chi_{g1} - \chi_{g2}$:
\begin{equation}
\label{polarised}
\ket{\psi_1}+e^{i\left(\chi_{g1} - \chi_{g2} \right)}\ket{\psi_2}\to
\ket{R}+e^{i\left(\chi_{g1} - \chi_{g2}\right)}\ket{L}\text{.}
\end{equation}
The detection of  of the emitted photon provides therefore a simple detection scheme of
the geometric phase.

\section{Conclusions}
 In the above sections we have described a novel scheme to generate geometric phases by cyclically modifying the irreversible dynamics of a quantum system. This can be seen as a parallel transport of a decoherence free subspace generated by a cyclic change of a rigged reservoir. The reservoir we consider is a broadband squeezed vacuum. The scheme can be discussed in more general terms, as shown in \cite{vlatko}. One can think to extend the above scheme to other specific scenario.

 Again we stress that the interesting feature of the phases so generated is that they are intrinsically immune to noise. Furthermore in the scheme analyzed above a specific detection procedure which does not need any spin echo has been proposed.

\section*{Acknowledgments}
This work was supported in part by the EU under grant IST - TOPQIP, "Topological Quantum Information Processing" (Contract
IST-2001-39215).

\end{document}